\documentclass[12pt]{article}
\usepackage{amsmath, amssymb}
\usepackage{graphicx}

\usepackage[super,sort&compress]{natbib}
\setcitestyle{comma}

\usepackage{url} 
\usepackage{color, colortbl}
\usepackage{geometry}
\usepackage{array}
\usepackage{caption}
\usepackage{subcaption}
\usepackage{setspace}
\usepackage{hyperref}
\hypersetup{
  colorlinks= true,
  linkcolor= cyan,
  citecolor= blue,
  urlcolor= blue
}
\usepackage{float}

\newcommand{\btheta}{\boldsymbol{\theta}}
\newcommand{\bbeta}{\boldsymbol{\beta}}

\newcommand{\bxi}{\boldsymbol{\xi}}
\newcommand{\bfeta}{\boldsymbol{\eta}}
\newcommand{\tilgamma}{\widetilde{\gamma}}
\newcommand{\tillambda}{\widetilde{\lambda}}

\definecolor{lightgray}{gray}{0.9} 

\let\oldbibliography\thebibliography
\renewcommand{\thebibliography}[1]{\oldbibliography{#1}
\setlength{\itemsep}{-1pt}}

\begin{document}

\title{\vspace{-1.5cm} \bf Laplacian-P-splines for shared Gamma frailty models applied to clustered right-censored time-to-event data}

\author{Piotr Lewczuk\textsuperscript{1,2,*},
Oswaldo Gressani\textsuperscript{3},
Steven Abrams\textsuperscript{3,4}, 
Christel Faes\textsuperscript{3}}
\date{}

\maketitle

{\footnotesize
\noindent \textsuperscript{1}Lab for Clinical Neurochemistry, Department of Neurology, Universitätsklinikum Erlangen, Germany; \\
\textsuperscript{2}Department of Neurodegeneration Diagnostics, University Hospital Bialystok, Poland; \\
\textsuperscript{3}Interuniversity Institute for Biostatistics and statistical Bioinformatics (I-BioStat), Data Science Institute, Hasselt University, Belgium; \\
\textsuperscript{4}Global Health Institute, Family Medicine and Population Health, University of Antwerp, Belgium. \\
\textsuperscript{*}Correspondence: Piotr.Lewczuk@uk-erlangen.de
\par}

\linespread{1}
\begin{abstract}
\addcontentsline{toc}{section}{Abstract}
\noindent Shared frailty models have been proposed to accommodate unmeasured cluster-specific risk factors through the inclusion of a common latent frailty term. Among possible frailty distributions, the Gamma distribution is appealing due to its non-negativity, flexibility, and algebraic tractability leading to closed-form marginal survival or hazard function expressions. Under the Bayesian paradigm, the posterior distributions of model parameters are usually explored with computationally intensive procedures relying on Markov chain Monte Carlo sampling. As an alternative, Laplacian-P-splines (LPS) provide a flexible and sampling-free alternative by relying on Gaussian approximations of the posterior target distributions. In this model class, analytical formulas are obtained for the gradient and Hessian, yielding a computationally efficient inference scheme for estimation of model parameters with a natural way of quantifying uncertainty. This article extends the LPS toolbox to the inclusion of shared Gamma frailty models for clustered time-to-event data. We assess the finite-sample performance of the LPS estimation procedure through an extensive simulation study and compare estimates with those obtained using penalized partial likelihood estimation, without specification of the baseline hazard, and with the variance of the frailty term being estimated using profile likelihood. Finally, the proposed LPS estimation method is exemplified using three publicly available biomedical datasets on: (i) recurrent infections in children, (ii) cancer prevention, and (iii) kidney transplantation. \\[8pt]
\noindent \textit{\textbf{Key words:} Laplace approximations; Bayesian P-splines; Shared Gamma frailty models; Survival analysis.}\\[8 pt]
\end{abstract}

\vspace{-1.5cm}\noindent\hrulefill

\section{Introduction}
Cox proportional hazards (CPH) models in survival analysis assume that observed covariates act multiplicatively on a baseline hazard function, typically through a function $r(\boldsymbol{x}) = \exp(\boldsymbol{\beta}^{\top}\boldsymbol{x})$, implying that event times are assumed to be conditionally independent given observed covariate information \cite{cox72}. However, unobserved (latent) heterogeneity through unmeasured covariates often exists in practice, leading to biased estimates when ignored; for example, Balan and Putter\cite{bapu} provide a practical illustration of the impact of unmeasured covariates. To account for this, models have been developed that incorporate latent random effects capturing unobserved risk factors and quantifying unobserved heterogeneity. Consider, for illustrative purpose, a multicenter study of time to adverse effect after a new surgical procedure. In this setting it is plausible that patients within a center share common latent characteristics due to being treated in this particular center which are different from characteristics in patients being treated elsewhere. Alternatively, we may be interested in time between placement of dental fillings and development of secondary caries. Then, again, times to development of secondary caries for different teeth of the same patient are likely dependent, due to the existence of patient-specific predispositions. A special scenario is recurrent appearance of the event of interest (say, time elapsed between episodes of migraine or seasonal flu) in a given subject. In line with Abrams \textit{et al.} \cite{Ab1}, it might be postulated that recurrently appearing life-threatening infections may steadily worsen overall patient's condition, leading to increased vulnerability of infections in the future or, in contrary, infection may induce long-term immunity, protecting patients from another infection with the same pathogen. Therefore, association between recurrent event times can be accounted for by adding a random frailty effect. In all these examples, the time-to-event data are characterized by a hierarchical structure due to the clustering of observations within the same individual (individual-level multivariate time-to-event data) or across individuals within the same cluster.

\par Although the concept of univariate ``frailty" to measure unobserved heterogeneity dates back to the early work by Beard \cite{beard_59}, a shared frailty was first introduced for the analysis of multivariate time-to-event data by Holt and Prentice \cite{Holt1974} and Clayton \cite{Clayton1978} to accommodate unobserved cluster-specific risk factors. Among possible distributions of the frailty random variable, the Gamma distribution, which is a special case of the Power Variance Function (PVF) family, is attractive since it provides a closed-form expression for the Laplace transform, which is particularly useful when integrating out the frailty term in the conditional survival function to obtain the marginal survival function. Alternatively, the population-averaged hazard function has an analytically tractable expression under the Gamma frailty distribution. An additional reason for considering this distribution as a starting point is the fact that it is the only continuous distribution having constant local association over time. The choice of the frailty distribution determines the underlying association structure of the cluster-specific event times and, as Hougaard noticed, Gamma models assume that the dependence is emphasized in late events \cite{houg}. Alternative distributions are also described in the literature, such as lognormal (without closed form Laplace transform, though closely related to the choice of normal random effects in mixed models), inverse Gaussian or PVF frailty distributions, with an overview of different possible choices detailed in Wienke \cite{wienke2010}. 

\par Estimating parameters in shared frailty models involves two major challenges, namely integrating out the frailty term to obtain the marginal likelihood, and estimating both the regression coefficients and the frailty variance. In a frequentist setting, maximum likelihood estimation is often used to estimate parameters of the marginal model, which typically involves numerical optimization. Another frequentist approach is to treat the frailty term as missing and to apply the Expectation-Maximization algorithm where, in the E-step, one calculates the expected complete-data likelihood given the current estimates, followed by (log-)likelihood maximization in the M-step. For Gamma frailty distributions, analytical forms of the E-step often exist via the Laplace transform of the Gamma distribution \cite{gamage}. Other ways of estimating the model parameters in (shared) frailty models are detailed in the literature \cite{bapu}. In contrast, classical Bayesian methods assign (hyper)priors to the parameters of interest and use Markov chain Monte Carlo (MCMC) algorithms to sample posterior distributions.

\par Laplacian-P-splines (LPS) have been shown to be a powerful tool for Bayesian inference for different model classes by offering a sampling-free alternative to MCMC. The flexibility of P-spline smoothers, combined with Laplace approximations to selected posterior target distributions has opened up a modelling path that delivers a fast and accurate method for Bayesian inference in (generalized) additive models \cite{gress2021}, epidemic models \cite{gress_2023, gress_tor, RN2280, sumal}, additive proportional odds models \cite{lamgress2023}, and survival models \cite{RN2283, Gress2018}. Key quantities associated to the Laplace approximation of a target distribution are the gradient and the Hessian matrix, as they are used in iterative algorithms to compute the mode of the posterior distribution related to the parameters of interest.

\par This paper extends the LPS toolbox to the case of shared Gamma frailty models for clustered time-to-event data under right censoring. More specifically, it combines Laplace approximations for fast inference and P-splines for flexible modeling of the baseline hazard function. The paper is organized as follows. Section 2 describes the Bayesian P-splines shared Gamma frailty model and the Laplace approximation scheme for sampling-free inference. Section 3 assesses the performance of the LPS method across different simulation scenarios and Section 4 highlights the application of the proposed LPS approach to three real-world biomedical datasets on: (i) recurrent infections in children, (ii) cancer prevention study, and (iii) kidney transplantation. Finally, in Section 5, we discuss avenues for further research and formulate a conclusion with regard to the findings presented in this paper.

\section{Bayesian shared Gamma frailty model with P-splines and Laplace approximations}
\vspace{-0.25cm}
\noindent We assume a random right censoring scheme, where we observe a realization of a random variable $T=\min(T^*,C)$, i.e.,\ the minimum of a non-negative random variable $T^*$ (the event time of interest) and a non-negative random censoring time $C$. Further, we assume that censoring is independent and non-informative, i.e., censoring times are independent of the event times, and parameters of the censoring distribution are distinct from the parameters governing the event time distribution. An event indicator $\Delta$ specifies which of $T^*$ or $C$ arises first, i.e.,\ $\delta = 1$ if $T^*\leq C$ and $\delta = 0$ otherwise. \\

\subsection{Gamma frailty model}
\indent We consider a dataset of $n$ observations subgrouped into $\mathcal{I}$ mutually exclusive clusters, such that the $i$th cluster consists of $n_i$ observations sharing the same frailty. By $i=1,\dots,\mathcal{I}$ and $j=1,\dots,n_i$, we index the $j$th observation in the $i$th cluster. We assume that the frailty term $U$ is an unobserved positive random variable with a probability density function (pdf) $f_U$. In order to ensure identifiability of the frailty model, we assume that the cluster-specific frailty term follows a Gamma distribution, $U \sim \mathcal{G}(\gamma, \gamma)$, with unit mean $\mathbb{E}(U)=1$ and frailty variance $\mathbb{V}(U)=1/\gamma$. Under this specification, the frailty is cluster-specific, such that the conditional hazard is the hazard which is shared among all subjects in the same cluster having the same covariate information. Thus, the conditional hazard for subject $j$ in cluster $i$ with an associated profile of covariates represented by the $p$-dimensional column vector $\boldsymbol{z}_{ij}=(z_{ij1},\dots,z_{ijp})^{\top}$ renders the form:

\vspace{-1cm}

\begin{eqnarray}
h_{ij}(t \vert u_i) = h_0(t) u_i \exp(\bbeta^{\top}\boldsymbol{z}_{ij}),
\label{cond_hazard}
\end{eqnarray}

\noindent where $u_i$, $i=1,\dots,\mathcal{I}$, is a realization of the shared frailty for all subjects in cluster~$i$, $h_0(\cdot)$ is the baseline hazard modelled with cubic B-splines $h_0(t)=\exp(\btheta^{\top}\boldsymbol{b}(t))$ with $\btheta=(\theta_1,\dots,\theta_K)^{\top}$, and $\boldsymbol{b}(\cdot)=(b_1(\cdot),\dots,b_K(\cdot))^{\top}$ a B-spline basis defined on the compact support $\mathcal{B}=[0,t_r]$, where $t_r$ is the largest observed follow-up time. The regression coefficients related to covariate effects are collected into the vector $\bbeta=(\beta_1,\dots,\beta_p)^{\top}$.

\indent Let $t_{ij}$ be the observed failure or censoring time for subject $j$ in cluster $i$, i.e.,\ $t_{ij}$ is a realization of $T_{ij}$ and $\delta_{ij}$ is a realization of the associated binary indicator, $\Delta$. As such, the information from the $j$th individual in cluster $i$ becomes $\mathcal{D}_{ij}=(t_{ij},\delta_{ij},\boldsymbol{z}^{\top}_{ij})$ and accounting for all individuals in all clusters, we denote the complete data by $\mathcal{D}$. Further, with $S_0$ and $H_0$ we define, respectively, the baseline survival and cumulative hazard functions, such that $H_0(t) = \int_0^{t} h_0(s) ds$ and $S_0(t) = \exp(-H_0(t))$. Then, the likelihood contribution to the $i$th cluster becomes \cite{wienke2010}:

\vspace{-1cm}

\begin{align}
\mathcal{L}_i(\btheta, \bbeta \vert u_i; \mathcal{D}_i)&= \prod_{j=1}^{n_i} \left(h_0(t_{ij}) u_i \exp(\bbeta^{\top} \boldsymbol{z}_{ij}) \right)^{\delta_{ij}} \exp\left(-H_0(t_{ij}) u_i \exp(\bbeta^{\top} \boldsymbol{z}_{ij}) \right) \nonumber \\
&\approx \prod_{j=1}^{n_i} \Biggl\{ \left(\exp(\btheta^{\top}\boldsymbol{b}(t_{ij}) + \bbeta^{\top} \boldsymbol{z}_{ij}) u_i  \right)^{\delta_{ij}} \nonumber  \\
&\times \exp\left(-\Bigg(\sum_{l=1}^{l(t_{ij})} \exp(\btheta^{\top}\boldsymbol{b}(s_l)) \Delta^* \Bigg) u_i \exp(\bbeta^{\top} \boldsymbol{z}_{ij}) \right)
\Biggr\}.
\label{lik}
\end{align}


\noindent The Riemann midpoint approximation in the last line follows from the fact that the cumulative hazard function has no analytic solution and the definite integral is approximated on a grid of equidistant points. By $l(t_{ij})$, we denote the grid segment containing the time point $t_{ij}$, $s_l$ denotes the midpoint of that segment, and $\Delta^*$ is the length of a segment. The marginal likelihood for cluster $i$ is obtained by integrating out the frailty variable $u_i$ in Eq.\ \eqref{lik}. In general, the marginal likelihood contribution within the $i$th cluster renders the form~\cite{DJA}:

\vspace{-1cm}

\begin{align}
\mathcal{L}_i(\boldsymbol{\theta}, \boldsymbol{\beta}; \mathcal{D}_i) &= 
\int_{0}^{+\infty}{\prod_{j = 1}^{n_i}{\Bigg\{u^{\delta_{ij}}\exp(\boldsymbol{\theta}^{\top}\boldsymbol{b}(t_{ij}) + \boldsymbol{\beta}^{\top}\boldsymbol{z}_{ij})^{\delta_{ij}}}} \nonumber \\
&\quad \times \left.\exp\left(-\left(\sum_{l = 1}^{l(t_{ij})}{\exp(\boldsymbol{\theta}^{\top} \boldsymbol{b}(s_l))\Delta^*}\right)u\exp(\boldsymbol{\beta}^{\top} \boldsymbol{z}_{ij})\right) \right\} f_U(u) \mbox{d}u \nonumber \\
&=
\left(-1\right)^{n_i\bar{\delta}_i} \mathbb{L}_U^{(n_i\bar{\delta}_i)} \left(\sum_{j = 1}^{n_i}\left\{\left(\sum_{l = 1}^{l(t_{ij})}{\exp(\boldsymbol{\theta}^{\top} \boldsymbol{b}(s_l))\Delta^*}\right)\exp(\boldsymbol{\beta}^{\top} \boldsymbol{z}_{ij})\right\}\right) \nonumber \\
&\quad \times
\prod_{j=1}^{n_i}{\left(h_0(t_{ij})\exp(\boldsymbol{\beta}^{\top} \boldsymbol{z}_{ij})\right)^{\delta_{ij}}},
\label{general}
\end{align}

\noindent where $n_i\bar{\delta}_i := \sum_{j = 1}^{n}{\delta_{ij}}$ represents the number of events in cluster $i$, $\mathbb{L}_U(\cdot)$ is the Laplace transform of the frailty random variable $U$, and $\mathbb{L}^{(k)}(\cdot)$ is the $k$th order derivative of $\mathbb{L}_U(\cdot)$. For the particular case of a Gamma frailty distribution with Laplace transform $\mathbb{L}_U(s) = \mathbb{E}(\exp(-sU)) = (1 + \gamma^{-1}s)^{-\gamma}$, the marginal likelihood for cluster $i$ becomes:

\vspace{-0.5cm}

\begin{eqnarray}
\label{integfrail1}
\mathcal{L}_{i}(\btheta,\bbeta, \gamma; \mathcal{D}_i)\hspace{-0.2cm}&=& \prod_{j=1}^{n_i} \exp(\btheta^{\top}\boldsymbol{b}(t_{ij})+\bbeta^{\top} \boldsymbol{z}_{ij})^{\delta_{ij}} \nonumber \\
&& \times \int_{0}^{+\infty} u^{\sum_{j=1}^{n_i} \delta_{ij}} \exp\left(-u \sum_{j=1}^{n_i}\left\{\Bigg(\sum_{l=1}^{l(t_{ij})} \exp(\btheta^{\top}\boldsymbol{b}(s_l)) \Delta^* \Bigg) \exp(\bbeta^{\top} \boldsymbol{z}_{ij}) \right\}\right)\nonumber \\
&& \times \frac{\gamma^{\gamma}}{\Gamma(\gamma)} u^{\gamma-1} \exp(-\gamma u) \ du \nonumber \\
&=& \frac{\gamma^{\gamma}}{\Gamma(\gamma)} \left(\prod_{j=1}^{n_i} \exp(\btheta^{\top}\boldsymbol{b}(t_{ij})+\bbeta^{\top} \boldsymbol{z}_{ij})^{\delta_{ij}}\right) \nonumber \\
&& \times \int_{0}^{+\infty} u^{n_i \bar{\delta}_i+\gamma-1} \exp\left(-u \left[ \sum_{j=1}^{n_i}\left\{\Bigg(\sum_{l=1}^{l(t_{ij})} \exp(\btheta^{\top}\boldsymbol{b}(s_l)) \Delta^* \Bigg) \exp(\bbeta^{\top} \boldsymbol{z}_{ij}) \right\} + \gamma \right] \right) \ du. \nonumber \\
&&
\end{eqnarray}

\noindent Note that the integrand in Eq.\ \eqref{integfrail1} is the kernel of a Gamma distribution with shape parameter

\vspace{-1cm}

\begin{eqnarray*}
a_{\mathcal{G}}&=&n_i \bar{\delta}_i+\gamma, \nonumber \\
\end{eqnarray*}

\vspace{-1cm}

\noindent and rate parameter
\begin{eqnarray*}
b_{\mathcal{G}}&=&\sum_{j=1}^{n_i}\left\{\Bigg(\sum_{l=1}^{l(t_{ij})} \exp(\btheta^{\top}\boldsymbol{b}(s_l)) \Delta^* \Bigg) \exp(\bbeta^{\top} \boldsymbol{z}_{ij})\right\} + \gamma. \nonumber 
\end{eqnarray*}
\\

\vspace{-0.9cm}

\noindent Therefore, we can rewrite the last expression to obtain:

\begin{eqnarray}
\label{integfrail2}
\mathcal{L}_{i}(\btheta, \bbeta, \gamma; \mathcal{D}_i)\hspace{-0.2cm}&=&\hspace{-0.2cm} \frac{\gamma^{\gamma}}{\Gamma(\gamma)} \left(\prod_{j=1}^{n_i} \exp(\btheta^{\top}\boldsymbol{b}(t_{ij})+\bbeta^{\top} \boldsymbol{z}_{ij})^{\delta_{ij}}\right) \int_{0}^{+\infty} u^{a_{\mathcal{G}}-1} \exp\left(-u b_{\mathcal{G}} \right) \ du \nonumber \\
\hspace{-0.2cm}&=&\hspace{-0.2cm} \frac{\gamma^{\gamma}}{\Gamma(\gamma)} \left(\prod_{j=1}^{n_i} \exp(\btheta^{\top}\boldsymbol{b}(t_{ij})+\bbeta^{\top} \boldsymbol{z}_{ij})^{\delta_{ij}}\right) \Gamma(a_{\mathcal{G}}) b_{\mathcal{G}}^{-a_{\mathcal{G}}},
\end{eqnarray}

\noindent from where the marginal log-likelihood of cluster $i$ is easily obtained:

\vspace{-1.2cm}

\begin{eqnarray}
\ell_i(\btheta,\bbeta, \gamma; \mathcal{D}_i)&:=&\log \mathcal{L}_i(\btheta,\bbeta, \gamma; \mathcal{D}_i) \nonumber \\
&=& \gamma \log(\gamma) + \log(\Gamma(n_i \bar{\delta}_i+\gamma))- \log(\Gamma(\gamma)) + \sum_{j=1}^{n_i}\delta_{ij}\left(\btheta^{\top}\boldsymbol{b}(t_{ij})+\bbeta^{\top} \boldsymbol{z}_{ij}\right) \nonumber \\
&&- (n_i \bar{\delta}_i+\gamma) \log\left(\sum_{j=1}^{n_i}\left\{\Bigg(\sum_{l=1}^{l(t_{ij})} \exp(\btheta^{\top}\boldsymbol{b}(s_l)) \Delta^* \Bigg) \exp(\bbeta^{\top} \boldsymbol{z}_{ij})\right\} + \gamma \right). \nonumber 
\end{eqnarray}

\noindent From here on, we decide to work with the frailty parameter re-expressed on the log-scale, denoted by $\tilgamma=\log(\gamma)$. 

\subsection{Bayesian model}
\noindent P-splines can be integrated into a Bayesian framework by introducing a stochastic version of the difference penalty \cite{LangBrez}. A diffuse prior on the initial spline coefficient is often assumed, followed by a random walk imposed on the subsequent coefficients. Further, the penalty matrix can be defined, resulting in a proper prior for the vector of the B-spline coefficients conditional on the penalty parameter \cite{BrezStein, gressthes}. Let us define the vector of latent variables (including all parameters to be estimated) by $\bxi=(\btheta^{\top}, \bbeta^{\top}, \tilgamma)^{\top}$ in $\mathbb{R}^{K+p+1}$. We impose a proper conditional Gaussian prior on the latent vector $(\boldsymbol{\bxi} \vert \lambda)\sim \mathcal{N}_{\text{dim}(\bxi)}(\boldsymbol{0},\Sigma_{\lambda})$ with block diagonal variance-covariance matrix $\Sigma_{\lambda}=\text{diag}\left((\lambda P)^{-1}, \zeta^{-1} I_{p+1} \right)$ and $K \times K$ penalty matrix $P=D_r^{\top}D_r + \epsilon I_K$ associated to the spline components, where adding a small positive perturbation, $\epsilon$, on the diagonal ensures that the prior is proper. The identity matrix $I_{p+1}$ multiplied by a scalar (here $\zeta=10^{-6}$) translates into a non-informative prior belief on the regression coefficients and the log-frailty term. Following Jullion and Lambert \cite{jullion2007robust}, a robust hierarchical prior is imposed on the penalty parameter $\lambda >0$, namely $\lambda \vert \kappa \sim \mathcal{G}(\nu/2,(\nu \kappa)/2)$ where $\kappa \sim \mathcal{G}(a_{\kappa}, b_{\kappa})$ with $a_{\kappa}=b_{\kappa}=10^{-4}$ and $\nu=3$. Finally, the vector of hyperparameters is denoted by $\boldsymbol{\eta}=(\lambda, \kappa)^{\top}$.

\subsection{Laplace approximation to the conditional posterior $p(\bxi \vert \bfeta, \mathcal{D})$}
Having specified the model, our first goal is to approximate the posterior density $p(\bxi \vert \lambda, \mathcal{D})$ by a Gaussian density $\widetilde{p}_{G}(\bxi \vert \lambda, \mathcal{D})$ through a Laplace approximation. From Bayes' rule, the posterior distribution becomes $p(\bxi \vert \lambda, \mathcal{D}) \propto \mathcal{L}(\bxi; \mathcal{D}) p(\bxi \vert \lambda)$ which, using the analytical formula for the prior, gives $p(\bxi \vert \lambda, \mathcal{D}) \propto \exp\left(\ell(\bxi; \mathcal{D})-0.5 \bxi^{\top} Q_{\lambda} \bxi \right)$, where $\ell (\cdot) = \sum_{i=1}^\mathcal{I} \ell_i(\cdot) $, with prior precision matrix $Q_{\lambda}:=\Sigma^{-1}_{\lambda}=\text{diag}(\lambda P, \zeta I_{p+1})$. The log conditional posterior is thus given by $\log p(\bxi \vert \lambda, \mathcal{D}) \dot{=} \ell(\bxi; \mathcal{D})-0.5 \bxi^{\top} Q_{\lambda} \bxi$, where $\dot{=}$ denotes equality up to an additive constant. A Laplace approximation, centered around the (unknown) mode $\bxi_{M}$ of $\log p(\bxi \vert \lambda, \mathcal{D})$, can be obtained through a Levenberg-Marquardt Algorithm (LMA). The latter makes use of the gradient 
$\nabla_{\bxi} \log p(\bxi \vert \lambda, \mathcal{D}) = \nabla_{\bxi} \ell(\bxi; \mathcal{D})-Q_{\lambda} \bxi$ and Hessian matrix $\nabla_{\bxi}^2 \log p(\bxi \vert \lambda, \mathcal{D}) = \nabla_{\bxi}^2 \ell(\bxi; \mathcal{D})-Q_{\lambda}$ (see \textbf{Appendix} for analytical derivations). At convergence of the algorithm, we recover the Gaussian approximation $p(\bxi \vert \lambda, \mathcal{D}) \leftarrow \widetilde{p}_{G}(\bxi \vert \lambda, \mathcal{D})=\mathcal{N}_{\text{dim}(\bxi)}(\bxi^{*}_{\lambda}, \Sigma^{*}_{\lambda})$, where $\bxi^*_{\lambda}$ is an approximation to the posterior mode $\bxi_{M}$ conditional on the penalty parameter $\lambda$ and $\Sigma^*_{\lambda}=-(\nabla_{\bxi}^2 \log p(\bxi \vert \lambda, \mathcal{D}))^{-1}\vert_{\bxi = \bxi^*_{\lambda}}$.

\subsection{Estimation of the hyperparameters $\bfeta$}
\noindent Next, we derive the approximation to the posterior distribution of the penalty parameter $\lambda$. The marginal posterior distribution of $\bfeta=(\lambda, \kappa)^{\top}$ can be written as follows:

\vspace{-0.5cm}

\begin{eqnarray}
p(\bfeta \vert \mathcal{D}) \propto \frac{\mathcal{L}(\bxi; \mathcal{D}) p(\bxi \vert \lambda) p(\lambda \vert \kappa) p(\kappa)}{p(\bxi \vert \lambda, \mathcal{D})}, \nonumber 
\end{eqnarray}

\noindent and by replacing the denominator with the Laplace approximation around $\bxi^*_{\lambda}$ and evaluating $\bxi$ at the latter value, we obtain an approximation to the marginal posterior distribution of the hyperparameter vector:

\vspace{-0.4cm}

\begin{eqnarray}
\widetilde{p}(\bfeta \vert \mathcal{D}) \propto \frac{\mathcal{L}(\bxi; \mathcal{D}) p(\bxi \vert \lambda) p(\lambda \vert \kappa) p(\kappa)}{\widetilde{p}_G(\bxi \vert \lambda, \mathcal{D})}\Bigg\vert_{\bxi = \bxi^*_{\lambda}}. \nonumber 
\end{eqnarray}

\noindent Writing the above expression in extensive form, we have:

\vspace{-0.7cm}

\begin{eqnarray*}
\widetilde{p}(\bfeta \vert \mathcal{D}) \propto  \exp\left(\ell(\bxi^*_{\lambda}; \mathcal{D})-0.5 \bxi^{*\top}_{\lambda}Q_{\lambda} \bxi^*_{\lambda}\right) \lambda^{0.5 (K+\nu)-1} \big\vert \Sigma^*_{\lambda} \big\vert^{0.5} \kappa^{0.5\nu + a_{\kappa}-1} \exp(-\kappa(0.5\lambda \nu +b_{\kappa})).
\end{eqnarray*}

\vspace{-0.3cm}

\noindent Analytical integration with respect to $\kappa$ is possible as we recognize in the above expression the kernel of a Gamma distribution for $\kappa$, so that:

\vspace{-1cm}

\begin{eqnarray}
\widetilde{p}(\lambda \vert \mathcal{D}) &=& \int_{0}^{+\infty} \widetilde{p}(\bfeta \vert \mathcal{D}) d\kappa \nonumber \\
&\propto& \exp\left(\ell(\bxi^*_{\lambda}; \mathcal{D})-0.5 \bxi^{*\top}_{\lambda}Q_{\lambda} \bxi^*_{\lambda}\right) \lambda^{0.5 (K+\nu)-1} \big\vert \Sigma^*_{\lambda} \big\vert^{0.5} (0.5\lambda \nu +b_{\kappa})^{-(0.5\nu + a_{\kappa})}. \nonumber 
\end{eqnarray}

\noindent For numerical reasons, we rewrite the above expression in log terms, i.e.\, $\tillambda=\log (\lambda)$; hence the log posterior (taking the Jacobian of this transformation, $\exp(\widetilde{\lambda})$, into account) becomes:

\vspace{-1cm}

\begin{eqnarray}
\log \widetilde{p}(\tillambda \vert \mathcal{D}) &\dot{=}& \ell(\bxi^*_{\tillambda}; \mathcal{D}) -0.5\bxi^{*\top}_{\tillambda}Q_{\tillambda} \bxi^*_{\tillambda} + 0.5 (K+\nu) \tillambda \nonumber \\ &+& 0.5 \log \big\vert \Sigma^*_{\tillambda} \big\vert-(0.5 \nu + a_{\kappa}) \log\left(0.5 \exp(\tillambda) \nu + b_{\kappa} \right). \nonumber
\end{eqnarray}

\noindent From here the \textit{maximum a posteriori} (MAP) estimate of the (log) penalty term, i.e.\, ${\tillambda^*=\text{argmax}_{\tillambda}\log \widetilde{p}(\tillambda \vert \mathcal{D})}$ can be computed. Plugging the latter into the Laplace approximation to the conditional posterior of the latent vector yields the final approximation, which can be used for posterior inference: 
\begin{equation}
\widetilde{p}_G(\boldsymbol{\xi} \vert \tillambda^*, \mathcal{D}) = \mathcal{N}_{\text{dim}(\boldsymbol{\xi})} \Bigl( \boldsymbol{\xi}^*_{\tillambda^*}, \; \boldsymbol{\Sigma}^*_{\tillambda^*} \Bigr).
\label{cond_post_xi_star}
\end{equation}

\subsection{Credible intervals for the baseline survival function}
Pointwise credible intervals (CrIs) for the baseline survival function, $S_0(t)$, can be computed using the Delta method and a $\log(-\log(\cdot))$ transformation \cite{Paw}. The latter transformation yields: $G(\btheta | t) = \log(-\log(S_0(t))) = \log(\sum_{l=1}^{l(t)} \exp(\btheta^{\top} \boldsymbol{b}(s_l)) \Delta^*) $. We denote the gradient of $G(\btheta|t)$ with respect to the vector of the spline coefficients by $\nabla_{\btheta} G(\btheta|t)$; its elements are:

\vspace{-0.1cm}
\begin{equation}
\frac{\partial G(\btheta | t)}{\partial \theta_k} = \frac{\sum_{l=1}^{l(t)} \exp(\btheta^{\top} \boldsymbol{b}(s_l)) b_k(s_l)}{\sum_{l=1}^{l(t)} \exp(\btheta^{\top} \boldsymbol{b}(s_l))},
\nonumber
\end{equation}
\\
\noindent for $k=1,\dots,K$. Then, a $(1 - \alpha) \times 100\%$ CrI for $G(\btheta|t)$ is:

\begin{equation}
\text{CrI}_{G(\btheta|t)} = G(\btheta^*|t) \pm z_{\alpha / 2} \Big[ \Big( \nabla_{\btheta}^{\top}G(\btheta|t)|_{\btheta=\btheta^*}   \Sigma^*_{\tillambda^*} \nabla_{\btheta}G(\btheta|t)|_{\btheta=\btheta^*} \Big) \Big]^{\frac{1}{2}},
\label{CrI}
\end{equation}
\\
\noindent where $z_{\alpha / 2}$ denotes $\alpha / 2$ quantile of the standard normal distribution, and $\nabla_{\btheta}^{\top}G(\btheta|t)|_{\btheta=\btheta^*}$ is the gradient evaluated at the vector of the spline coefficients resulting from the Laplace approximation in Eq.\ \eqref{cond_post_xi_star}. By back-transformation of the expression in Eq.\ \eqref{CrI} using a $\exp(-\exp(\cdot))$ transform, we obtain the desired pointwise CrI for the baseline survival function at time $t$.

\section{Simulation procedure and results}
In order to compare the LPS approach with the frequentist approach, a simulation study was performed.

\subsection{Simulation procedure}
\noindent Given the conditional hazard in the shared Gamma frailty model in Eq.\ \eqref{cond_hazard} and exploiting the exponential relation between the survival and the cumulative hazard functions, we obtain the conditional survival function:

\vspace{-1cm}

\begin{eqnarray}
	S_{ij}(t\vert u_i)&=&\exp\left(-\int_{0}^t h_{ij}(s \vert u_i)\ ds \right) \nonumber \\
	&=& \exp\left(- u_i \exp(\bbeta^{\top}\boldsymbol{z}_{ij}) H_0(t) \right). \nonumber
\end{eqnarray}

\noindent Hence, the cumulative distribution function of subject $j$ in cluster $i$ conditional on the realized cluster frailty $u_i$ becomes:

\vspace{-1cm}

\begin{eqnarray}
	\label{cdfij}
	F_{ij}(t \vert u_i)&=&1-S_{ij}(t \vert u_i) \nonumber \\
	&=& 1-\exp\left(- u_i \exp(\bbeta^{\top}\boldsymbol{z}_{ij}) H_0(t) \right). \nonumber
\end{eqnarray}

\noindent The generalized inverse of $F_{ij}$, conditional on $u_i$ (quantile function) can be defined as a function of $x$ with domain on the open interval $x \in (0,1)$ as follows ${F_{ij}^{-1}(x\vert u_i) := \inf\{t : F_{ij}(t \vert u_i) \geq x\}}$. If the quantile function is available in closed form, the inverse transform sampling method can be used to sample $T_{ij}$ from $F_{ij}(\cdot \vert u_i)$. In our frailty model setting, the inverse cumulative distribution function can be expressed as the inverse baseline cumulative hazard function:

\vspace{-1cm}

\begin{eqnarray}
	\label{invF}
	F_{ij}^{-1}(x \vert u_i) = H_0^{-1}\left(-\frac{\log(1-x)}{u_i \exp(\bbeta^{\top}\boldsymbol{z}_{ij})}\right).
\end{eqnarray}

\noindent From the inverse transformation method, if $X\sim \text{Unif}(0,1)$ (where $\text{Unif}(0,1)$ is a uniform distribution on the open interval $(0,1)$), then $T_{ij}=F_{ij}^{-1}(X \vert u_i) \sim F_{ij}(\cdot \vert u_i)$. We assume that the baseline hazard function corresponds to the hazard of a Weibull distribution with shape parameter $a>0$ and scale parameter $b>0$, and the pdf $f_0(t)=(a/b^a) t^{(a-1)}\exp(-(t/b)^a)$. The baseline survival function is then:

\vspace{-1cm}

\begin{eqnarray}
	S_0(t) =  \exp\left(-\left(\frac{t}{b}\right)^a\right) \nonumber,
\end{eqnarray}

\noindent and it follows that $H_0(t)=-\log S_0(t) = (t/b)^a$ and $H_0^{-1}(w)=bw^{1/a}$. The baseline hazard becomes $h_0(t)=(a/b^a) t^{(a-1)}$, and the quantile function in \eqref{invF} becomes:

\vspace{-1cm}

\begin{eqnarray}
	F_{ij}^{-1}(x \vert u_i) = b \left(-\frac{\log(1-x)}{u_i \exp(\bbeta^{\top}\boldsymbol{z}_{ij})}\right)^{\frac{1}{a}}.
\end{eqnarray}

\noindent We assume that the censoring time follows an exponential distribution with rate parameter $\varrho >0$, i.e.\, $C_{ij}\sim \text{Exp}(\varrho)\ \forall j \in \{1,...,n_i\}$ and denote by ${F_{ij}^C(c)=1-\exp(-\varrho c)}$ the cumulative distribution function of the censoring time evaluated in $c\geq 0$. Further, let $p_{ij}:=P(T_{ij} \leq C_{ij})$ denote the probability that the survival time is weakly smaller than the censoring time, so that the event indicator can be written as a Bernoulli random variable $\Delta_{ij} \sim \text{Bern}(p_{ij})$ with $\mathbb{E}(\delta_{ij})=p_{ij}$. With this notation, the expected censoring rate is given by:

\vspace{-1cm}

\begin{eqnarray}
	\mathbb{E}(\pi_{\text{cens}})&=&\mathbb{E}\left( \frac{1}{n} \sum_{i=1}^{\mathcal{I}}\sum_{j=1}^{n_i} \mathbb{I}(T_{ij} > C_{ij})\right) \nonumber \\
	&=&1-\frac{1}{n} \sum_{i=1}^{\mathcal{I}}\sum_{j=1}^{n_i} \exp(-\varrho T_{ij}).
\end{eqnarray}

\noindent Hence, a desired censoring rate $\mathbb{E}^*(\pi_{\text{cens}}) \in (0,1)$ can be achieved by finding the value of $\varrho^{\star}$ satisfying $1-\mathbb{E}^*(\pi_{\text{cens}})-\frac{1}{n} \sum_{i=1}^{\mathcal{I}}\sum_{j=1}^{n_i} \exp(-\varrho^{\star} T_{ij})=0$. This problem can be easily solved with a root finding algorithm. The pseudo-code below summarizes the data generation mechanism to obtain a full simulated dataset for the frailty model.\\

\vspace{0.5cm}

\hrule
\vspace{0.1cm}
\noindent {\textbf{Data generating mechanism for simulation study} \label{DGM}
\vspace{0.1cm}
\hrule 
\vspace{0.1cm}
\noindent 1:\ \text{Fix}\ $a,b>0$ (Weibull parameters); $\gamma>0$ (frailty term);  $\mathbb{E}^*(\pi_{\text{cens}}) \in (0,1)$ (expected censoring rate); $\mathcal{I} \in \mathbb{N}^*$ (number of groups); $n_i>0,\ i=1,\dots,\mathcal{I}$ and $\boldsymbol{\beta} \in \mathbb{R}^p$. \\
2:\ \text{\textbf{for} $i=1,\dots, \mathcal{I}$\ \textbf{do}}\\
3:\hspace{0.2cm} \text{Draw $u$ from } $\mathcal{G}(\gamma, \gamma)$\\
4:\hspace{0.7cm} \text{\textbf{for} $j=1,\dots, n_i$\ \textbf{do}}\\
5:\hspace{1.2cm} \text{Generate vector} $\boldsymbol{z}_{ij}$ using known parametric families\\
6:\hspace{1.2cm} \text{Draw $x$ from Unif}$(0,1)$ \\
7:\hspace{1.2cm} \text{Compute} $t_{ij}=b \left(-\frac{\log(1-x)}{u \exp(\bbeta^{\top}\boldsymbol{z}_{ij})}\right)^{\frac{1}{a}}$ \\
8:\hspace{0.6cm} \text{\textbf{end for}}\\
9:\ \text{\textbf{end for}}\\
10: Find $\varrho^{\star}$ satisfying $1-\mathbb{E}^*(\pi_{\text{cens}})-\frac{1}{n} \sum_{i=1}^{\mathcal{I}}\sum_{j=1}^{n_i} \exp(-\varrho^{\star} T_{ij})=0$\ \text{with}\ $n=n_1+\dots+n_{\mathcal{I}}$\\
11:\ \text{\textbf{for} $i=1,\dots, \mathcal{I}$\ \textbf{do}}\\
12:\hspace{0.7cm} \text{\textbf{for} $j=1,\dots, n_i$\ \textbf{do}}\\
13:\hspace{1.2cm} \text{Draw $c_{ij}$ from Exp}$(\varrho^{\star})$ \\
14:\hspace{1.2cm} \text{Compute} $\delta_{ij}=\mathbb{I}(t_{ij}\leq c_{ij})$\\
15:\hspace{0.7cm} \text{\textbf{end for}}\\
16:\ \text{\textbf{end for}}\\
17:\ \text{Return} $t_{ij}, \delta_{ij}, \boldsymbol{z}_{ij}$ for $i=1,\dots,\mathcal{I}$ and $j=1,\dots,n_i$.
\vspace{0.1cm}
\hrule

\vspace{1cm}

\subsection{Simulation results}
The data generating mechanism was used to simulate $S^* = 300$ datasets for which the parameters of the shared Gamma frailty model were estimated using the LPS method detailed in the previous section and implemented in R (version 4.3.2) \cite{gress_fra}. The LPS frailty routine showed a stable behaviour across all scenarios. In some scenarios, there was a negligible number of generated datasets for which the routine crashed due to optimization related issues. In that case, we simply generated another dataset according to the corresponding scenario to keep an automated and replicable simulation environment. Next to the LPS approach, we considered a penalized partial likelihood estimation approach in which the baseline hazard function is left unspecified (through the coxph function in the \textit{survival} package). Survival times were sampled from a Weibull distribution with shape and rate parameters equal to 5 and 70, respectively, and with the frailty parameter $\gamma$ set to 1.5. For each observation, values of one binary  and one continuous covariate were generated with those following a Bernoulli distribution (with probability 0.5) and standard normal distribution, respectively. Baseline survival was estimated using 15 B-spline basis functions with a second-order penalty on a grid of 300 equidistant segments covering $[0, \; t_{r,s}]$, for $s = 1,\dots,S^*$, where $t_{r,s}$ was the largest observation time for a given simulated dataset. Ten simulation scenarios with different sample sizes were considered, covering different combinations for the number of clusters ($10 - 50$), the number of observations in a cluster ($6 - 50$) and censoring percentages ($10 \% \text{ or } 20 \%$). In the simulation study, datasets were chosen to be balanced, i.e.\, all clusters in a given dataset consisted of the same number of observations. The empirical bias of the regression coefficients was calculated as the average of the difference between the estimate and its true value across all $S^*$ simulations. The empirical standard error (ESE) and root mean squared error (RMSE) are also provided. The coverage probability (CP) was computed based on 90\% and 95\% CrIs constructed for the LPS method and corresponding confidence intervals (CIs) related to the frequentist approach.
 
\par Tables \ref{tab_sim1} and \ref{tab_sim2} summarize the simulation results. Comparing results obtained by the two procedures clearly indicates that the estimates provided by the LPS algorithm are well in alignment with the benchmark frequentist competitor. Bias of the regression coefficients is, for both methods, about $5 \% - 10\%$ (in absolute terms) and somehow larger for the frailty parameter (though comparable between the two estimation methods). There is no obvious dependency of the bias (in either method) on the cluster sizes, number of observations in a cluster or censoring percentages. Similarly, ESE and RMSE are comparable in both methods and do not show systematic dependencies on the simulation scenarios. Also $90 \%$ and $95 \%$ CPs of the regression coefficients are comparable between the two procedures. LPS directly provides CPs for the $\gamma$ parameter, which are also well in plausible ranges.

\begin{table}[htb!]
	\caption{Scenarios 1--5 show simulation results for different combinations of cluster sizes and number of subjects per cluster under a censoring 		
	percentage of $10\%$. Results are reported for \textit{lpsfrail} and \textit{coxph} based on $S^*=300$ simulated datasets.}
	\setlength{\tabcolsep}{7pt} 
	\renewcommand{\arraystretch}{1.1} 
	\begin{tabular}{ccccccccccc}
	    \hline 
		\multicolumn{1}{l}{Scenario 1} & \multicolumn{1}{l}{$\mathcal{I}$} & \multicolumn{1}{l}{$n_i$} & \multicolumn{1}{l}{$n$} & \multicolumn{1}{l}{Parameter} & \multicolumn{1}{l}{Mean} & \multicolumn{1}{l}{Bias} & 
		\multicolumn{1}{l}{ESE} & 
		\multicolumn{1}{l}{RMSE} & 
		\multicolumn{1}{l}{$90\%$CP} & \multicolumn{1}{l}{$95\%$CP} \\ \hline
		& & & & $\beta_1=\phantom{-}0.693$ & 0.676 & -0.017 & 0.184 & 0.185 & 84.333 & 90.667 \\   
	    lpsfrail & 20 & 10 & 200 & $\beta_2=-0.150$ & -0.144 & 0.006 & 0.088 & 0.088 & 88.333 & 92.333\\  
	    & & & & $\gamma=\phantom{-}1.500$ & 1.927 & 0.427 & 0.756 & 0.867 & 89.333 & 96.333\\  
	    \hline   
	    & & & & $\beta_1=\phantom{-}0.693$ & 0.664 & -0.029 & 0.179 & 0.181 & 84.667 & 91.667\\   
	    coxph & 20 & 10 & 200 & $\beta_2=-0.150$ & -0.142 & 0.008 & 0.086 & 0.086 & 88.000 & 92.000\\  
	    & & & & $\gamma=\phantom{-}1.500$ & 1.991 &  0.491 & 0.816 & 0.951 &    NA  &   NA \\  
	    \hline 
	    \multicolumn{1}{l}{Scenario 2} & \multicolumn{1}{l}{$\mathcal{I}$} & \multicolumn{1}{l}{$n_i$} & \multicolumn{1}{l}{$n$} & \multicolumn{1}{l}{Parameter} & \multicolumn{1}{l}{Mean} & \multicolumn{1}{l}{Bias} & 
	    \multicolumn{1}{l}{ESE} & 
	    \multicolumn{1}{l}{RMSE} & 
	    \multicolumn{1}{l}{$90\%$CP} & \multicolumn{1}{l}{$95\%$CP} \\ \hline
	    & & & & $\beta_1=\phantom{-}0.693$ & 0.649 &  -0.044 & 0.076 & 0.087   & 82.000 & 89.000\\   
	    lpsfrail & 20 & 50 & 1000 & $\beta_2=-0.150$ & -0.140 & 0.010 & 0.035 & 0.036 & 88.000 & 94.000\\  
	    & & & & $\gamma=\phantom{-}1.500$ & 1.878 & 0.378 & 0.609 & 0.716 &    89.000 & 93.667\\  
	    \hline   
	    & & & & $\beta_1=\phantom{-}0.693$ & 0.661 &  -0.032 & 0.077 & 0.083 & 85.000 & 91.000\\   
	    coxph & 20 & 50 & 1000 & $\beta_2=-0.150$ & -0.143 & 0.007 & 0.036 & 0.036 & 88.667 & 94.333\\  
	    & & & & $\gamma=\phantom{-}1.500$ & 1.190 & -0.310 & 0.216 & 0.377 &   NA & NA \\ 
	    \hline 
	    \multicolumn{1}{l}{Scenario 3} & \multicolumn{1}{l}{$\mathcal{I}$} & \multicolumn{1}{l}{$n_i$} & \multicolumn{1}{l}{$n$} & \multicolumn{1}{l}{Parameter} & \multicolumn{1}{l}{Mean} & \multicolumn{1}{l}{Bias} & 
	    \multicolumn{1}{l}{ESE} & 
	    \multicolumn{1}{l}{RMSE} & 
	    \multicolumn{1}{l}{$90\%$CP} & \multicolumn{1}{l}{$95\%$CP} \\ \hline
	    & & & & $\beta_1=\phantom{-}0.693$ &  0.661 & -0.032 & 0.149 & 0.152 & 87.000 & 91.333\\   
	    lpsfrail & 50 & 6 & 300 & $\beta_2=-0.150$ & -0.140 & 0.010 & 0.080 & 0.080 & 85.667 & 92.000 \\  
	    & & & & $\gamma=\phantom{-}1.500$ & 1.783 & 0.283 & 0.441 & 0.523 & 89.667 & 96.000\\  
	    \hline   
	    & & & & $\beta_1=\phantom{-}0.693$ &  0.659 & -0.034 & 0.148 & 0.152 & 86.667  & 91.667\\   
	    coxph & 50 & 6 & 300 & $\beta_2=-0.150$ & -0.139 & 0.011 & 0.079 & 0.080  & 85.333 & 92.000 \\  
	    & & & & $\gamma=\phantom{-}1.500$ & 1.808 & 0.308 & 0.459 & 0.552 &    NA  &  NA\\ 
	    \hline 
	    \multicolumn{1}{l}{Scenario 4} & \multicolumn{1}{l}{$\mathcal{I}$} & \multicolumn{1}{l}{$n_i$} & \multicolumn{1}{l}{$n$} & \multicolumn{1}{l}{Parameter} & \multicolumn{1}{l}{Mean} & \multicolumn{1}{l}{Bias} & 
	    \multicolumn{1}{l}{ESE} & 
	    \multicolumn{1}{l}{RMSE} & 
	    \multicolumn{1}{l}{$90\%$CP} & \multicolumn{1}{l}{$95\%$CP} \\ \hline
	    & & & & $\beta_1=\phantom{-}0.693$ & 0.653 & -0.040 & 0.071 & 0.081 & 86.667 & 92.667 \\   
	    lpsfrail & 50 & 20 & 1000 & $\beta_2=-0.150$ & -0.138 & 0.012 & 0.036 & 0.038 & 87.000 & 93.333 \\  
	    & & & & $\gamma=\phantom{-}1.500$ &  1.817 & 0.317 & 0.366 & 0.483 & 83.000 & 91.667 \\  
	    \hline   
	    & & & & $\beta_1=\phantom{-}0.693$ & 0.666 & -0.027 & 0.072 & 0.077 &   88.000 &   95.000 \\   
	    coxph & 50 & 20 & 1000 & $\beta_2=-0.150$ & -0.140 & 0.010 & 0.037 & 0.038  &  87.000   & 94.000 \\  
	    & & & & $\gamma=\phantom{-}1.500$ & 1.482 & -0.018 & 0.266 & 0.267 &    NA  &  NA \\ 
	    \hline 
	    \multicolumn{1}{l}{Scenario 5} & \multicolumn{1}{l}{$\mathcal{I}$} & \multicolumn{1}{l}{$n_i$} & \multicolumn{1}{l}{$n$} & \multicolumn{1}{l}{Parameter} & \multicolumn{1}{l}{Mean} & \multicolumn{1}{l}{Bias} & 
	    \multicolumn{1}{l}{ESE} & 
	    \multicolumn{1}{l}{RMSE} & 
	    \multicolumn{1}{l}{$90\%$CP} & \multicolumn{1}{l}{$95\%$CP} \\ \hline
	    & & & & $\beta_1=\phantom{-}0.693$ & 0.688 & -0.005 & 0.180 & 0.180 & 87.333 & 92.667\\   
	    lpsfrail & 10 & 20 & 200 & $\beta_2=-0.150$ & -0.150 & 0.000 & 0.083 & 0.083 & 88.667 & 95.000 \\  
	    & & & & $\gamma=\phantom{-}1.500$ & 2.127 & 0.627 & 1.313 & 1.453 & 90.333 & 95.667   \\  
	    \hline   
	    & & & & $\beta_1=\phantom{-}0.693$ & 0.681 & -0.012 & 0.177 & 0.177 & 89.000 & 93.333\\   
	    coxph & 10 & 20 & 200 & $\beta_2=-0.150$ &  -0.148 & 0.002 & 0.083 & 0.083 & 89.667 & 95.000  \\  
	    & & & & $\gamma=\phantom{-}1.500$ &  1.967 & 0.467 & 1.386 & 1.461   &  NA   &  NA \\ 
	    \hline 
	\end{tabular}
\label{tab_sim1}
\end{table}

\begin{table}[htb!]
	\caption{Scenarios 6--10 show simulation results for different combinations of cluster sizes and number of subjects per cluster under a censoring 		
	percentage of $20\%$. Results are reported for \textit{lpsfrail} and \textit{coxph} based on $S^*=300$ simulated datasets.}
	\setlength{\tabcolsep}{7pt} 
	\renewcommand{\arraystretch}{1.1} 
	\begin{tabular}{ccccccccccc}
		\hline 
		\multicolumn{1}{l}{Scenario 6} & \multicolumn{1}{l}{$\mathcal{I}$} & \multicolumn{1}{l}{$n_i$} & \multicolumn{1}{l}{$n$} & \multicolumn{1}{l}{Parameter} & \multicolumn{1}{l}{Mean} & \multicolumn{1}{l}{Bias} & 
		\multicolumn{1}{l}{ESE} & 
		\multicolumn{1}{l}{RMSE} & 
		\multicolumn{1}{l}{$90\%$CP} & \multicolumn{1}{l}{$95\%$CP} \\ \hline
		& & & & $\beta_1=\phantom{-}0.693$ & 0.676 &  -0.017 & 0.198 & 0.199 & 84.333 & 90.667\\   
		lpsfrail & 20 & 10 & 200 & $\beta_2=-0.150$ & -0.150 & 0.000 & 0.092 & 0.092 & 87.000 & 93.333\\  
		& & & & $\gamma=\phantom{-}1.500$ & 1.961 & 0.461 & 0.812 & 0.932 & 89.333 & 96.000 \\  
		\hline   
		& & & & $\beta_1=\phantom{-}0.693$ & 0.663 & -0.030 & 0.192 & 0.194 & 86.000 & 91.667\\   
		coxph & 20 & 10 & 200 & $\beta_2=-0.150$ & 0.148 & 0.002 & 0.090 & 0.090 & 88.333 & 94.000 \\  
		& & & & $\gamma=\phantom{-}1.500$ & 2.050 & 0.550 & 0.875 & 1.032 &    NA  & NA \\  
		\hline 
		\multicolumn{1}{l}{Scenario 7} & \multicolumn{1}{l}{$\mathcal{I}$} & \multicolumn{1}{l}{$n_i$} & \multicolumn{1}{l}{$n$} & \multicolumn{1}{l}{Parameter} & \multicolumn{1}{l}{Mean} & \multicolumn{1}{l}{Bias} & 
		\multicolumn{1}{l}{ESE} & 
		\multicolumn{1}{l}{RMSE} & 
		\multicolumn{1}{l}{$90\%$CP} & \multicolumn{1}{l}{$95\%$CP} \\ \hline
		& & & & $\beta_1=\phantom{-}0.693$ & 0.647 & -0.046 & 0.080 & 0.092 & 81.667 & 89.000\\   
		lpsfrail & 20 & 50 & 1000 & $\beta_2=-0.150$ & -0.141 & 0.009 & 0.037 & 0.038 & 88.000 & 92.000 \\  
		& & & & $\gamma=\phantom{-}1.500$ & 1.884 &  0.384 &  0.621 & 0.729 & 89.000 & 93.667\\  
		\hline   
		& & & & $\beta_1=\phantom{-}0.693$ & 0.659 & -0.034  & 0.081 & 0.087 & 84.667 & 91.000\\   
		coxph & 20 & 50 & 1000 & $\beta_2=-0.150$ & -0.143 & 0.007 & 0.038 & 0.038 & 88.667 & 92.667 \\  
		& & & & $\gamma=\phantom{-}1.500$ &  1.225 & -0.275 & 0.257 & 0.376   &  NA   &  NA \\ 
		\hline 
		\multicolumn{1}{l}{Scenario 8} & \multicolumn{1}{l}{$\mathcal{I}$} & \multicolumn{1}{l}{$n_i$} & \multicolumn{1}{l}{$n$} & \multicolumn{1}{l}{Parameter} & \multicolumn{1}{l}{Mean} & \multicolumn{1}{l}{Bias} & 
		\multicolumn{1}{l}{ESE} & 
		\multicolumn{1}{l}{RMSE} & 
		\multicolumn{1}{l}{$90\%$CP} & \multicolumn{1}{l}{$95\%$CP} \\ \hline
		& & & & $\beta_1=\phantom{-}0.693$ & 0.659 & -0.034 & 0.158 & 0.162 & 86.000 & 92.667 \\   
		lpsfrail & 50 & 6 & 300 & $\beta_2=-0.150$ & -0.138 & 0.012 & 0.081 & 0.081 & 87.000 & 91.000  \\  
		& & & & $\gamma=\phantom{-}1.500$ & 1.819 & 0.319 & 0.490 & 0.584 & 90.333 & 95.333 \\  
		\hline   
		& & & & $\beta_1=\phantom{-}0.693$ & 0.655 & -0.038 & 0.157 & 0.162 & 85.333 & 92.667 \\   
		coxph & 50 & 6 & 300 & $\beta_2=-0.150$ & -0.138 & 0.012 & 0.080 & 0.080 & 87.000 & 90.667\\  
		& & & & $\gamma=\phantom{-}1.500$ & 1.865 & 0.365 & 0.521 & 0.635 &    NA  &   NA\\ 
		\hline 
		\multicolumn{1}{l}{Scenario 9} & \multicolumn{1}{l}{$\mathcal{I}$} & \multicolumn{1}{l}{$n_i$} & \multicolumn{1}{l}{$n$} & \multicolumn{1}{l}{Parameter} & \multicolumn{1}{l}{Mean} & \multicolumn{1}{l}{Bias} & 
		\multicolumn{1}{l}{ESE} & 
		\multicolumn{1}{l}{RMSE} & 
		\multicolumn{1}{l}{$90\%$CP} & \multicolumn{1}{l}{$95\%$CP} \\ \hline
		& & & & $\beta_1=\phantom{-}0.693$ & 0.655 & -0.039 & 0.076 & 0.085 & 89.000 & 92.667\\   
		lpsfrail & 50 & 20 & 1000 & $\beta_2=-0.150$ & -0.138 & 0.012 & 0.038 & 0.040 & 86.000 & 93.667 \\  
		& & & & $\gamma=\phantom{-}1.500$ & 1.819 & 0.319 & 0.371 & 0.488 & 84.667 & 92.000\\  
		\hline   
		& & & & $\beta_1=\phantom{-}0.693$ & 0.667 & -0.026 & 0.077 & 0.082 &  88.000 & 94.333 \\   
		coxph & 50 & 20 & 1000 & $\beta_2=-0.150$ & -0.140 & 0.010 & 0.039 & 0.040  &  87.000 & 94.000\\  
		& & & & $\gamma=\phantom{-}1.500$ &  1.536 & 0.036 & 0.298 & 0.300 &    NA  &   NA\\ 
		\hline 
		\multicolumn{1}{l}{Scenario 10} & \multicolumn{1}{l}{$\mathcal{I}$} & \multicolumn{1}{l}{$n_i$} & \multicolumn{1}{l}{$n$} & \multicolumn{1}{l}{Parameter} & \multicolumn{1}{l}{Mean} & \multicolumn{1}{l}{Bias} & 
		\multicolumn{1}{l}{ESE} & 
		\multicolumn{1}{l}{RMSE} & 
		\multicolumn{1}{l}{$90\%$CP} & \multicolumn{1}{l}{$95\%$CP} \\ \hline
		& & & & $\beta_1=\phantom{-}0.693$ & 0.690 & -0.003 & 0.191 & 0.191 & 88.000 & 93.333\\   
		lpsfrail & 10 & 20 & 200 & $\beta_2=-0.150$ & -0.148 & 0.002 & 0.088 & 0.088 & 90.000 & 94.667\\  
		& & & & $\gamma=\phantom{-}1.500$ & 2.234 &  0.734 & 1.686 & 1.837 & 89.667 & 94.333  \\  
		\hline   
		& & & & $\beta_1=\phantom{-}0.693$ & 0.681 & -0.012 & 0.187 & 0.187 & 89.333 & 94.333 \\   
		coxph & 10 & 20 & 200 & $\beta_2=-0.150$ & -0.145 & 0.005 & 0.087 & 0.087 & 89.333 & 95.667  \\  
		& & & & $\gamma=\phantom{-}1.500$ & 2.171 & 0.671 & 1.914 & 2.026 &    NA & NA \\ 
		\hline 
	\end{tabular}
\label{tab_sim2}
\end{table}

\section{Data applications}
The proposed LPS methodology was next applied to three publicly available datasets analyzed in previous studies. In all three datasets, the baseline hazard function was estimated with $K=30$ cubic B-splines and a second-order penalty. 

\subsection{CGD Study}
\par First, we readdress the data from a controlled randomized clinical trial on the application of Interferon Gamma (IG) to prevent serious, recurrent infections in children and young adults with Chronic Granulomatous Disease (CGD) \cite{RN2284}. The patient-level raw data are available in the literature \cite{frailtyEM, FlHarr} and the protocol of the trial is detailed elsewhere \cite{RN2284}. A total of 128 patients were randomized into an active arm ($n=63$, mean age $14.3 \pm 11.1$ years, $81\%$ males) treated with IG, and a control arm ($n=65$, mean age $15.0 \pm 9.6$ years, $82\%$ males) treated with placebo. Initially, the study was designed to consider the time, measured in days, between the randomization and the incidence of the first serious infection as the primary outcome. During the study, however, it turned out that several patients in both arms developed more than one occurrence; in such cases, multiple outcome variables per patient were recorded, namely the number of days between the randomization and the first infection and then between the end of a previous infection and the beginning of the next. This obviously leads to a hierarchical structure in the data, with observations clustered within patients and cluster sizes varying between 1 and 8. Approximately $63\%$ of the observations were right-censored. The main scientific interest focuses on the question whether the treatment with subcutaneous IG injections alters the hazard of serious infections, compared to the placebo treatment. Despite the fact that the cluster sizes are informative in this study, they reflect differences in event intensity across clusters, induced by potential treatment differences and cluster-specific variation, the latter being directly captured by the imposed frailty terms. In the current study, similarly to the original report, we considered two covariates, the indicator of the treatment arm (1 if IG and 0 otherwise) and sex (1 if female and 0 otherwise).

\par Results of the analyses with the LPS and the CPH approaches, presented in Table \ref{res_cgd}, indicate that the IG treatment reduces the hazard of serous infection in CGD subjects by a factor 3, i.e.\, the hazard of serious infection in the treatment group is 0.324 [0.165; 0.637] times the hazard of infection among patients in the placebo group. Sex does not have a substantial effect with the constant hazard ratio being not significantly different from one. The same conclusion, obtained in the original study, led to the introduction of IG for CGD treatment in 1990s, which meanwhile became standard. An advantage of the LPS approach over the CPH model is that it provides a smooth and fully modelled baseline survival curve, facilitating interpretation. The estimated survival curves resulting from the LPS model (Figure \ref{fig:rl_curves}, panel A) align well to the nonparametric survival curves obtained from the Kaplan-Meier (KM) estimator.

\vspace{0.5cm}

\begin{table}[htb!]
    \begin{center}
	\caption{Results of the LPS and the CPH in the CGD Study. CrI, CI, and IG denote Credible Interval, Confidence Interval and Interferon Gamma, 
		respectively.}
		\begin{tabular}{l | cccc}
			\hline
			\hline
            \multicolumn{1}{c |}{} & \multicolumn{2}{c}{LPS approach} & \multicolumn{2}{c}{CPH approach} \\
			Parameter & Estimate & $95\%$ CrI & Estimate & $95\%$ CI \\ 
			\hline
            IG Treatment     & $-1.127$  & [$-1.802$; $-0.451$] & $-1.132$ &  [$-1.793$; $-0.472$]  \\
            Female sex       & $-0.246$  & [$-1.010$; $\phantom{-}0.608$]    & $-0.232$ &  [$-1.082$;  $\phantom{-}0.619$] \\			
			$\gamma$         & $\phantom{-}0.698$     & $\phantom{-}[0.297; \phantom{-}1.638$]       & $\phantom{-}0.732$    &  -- \\
            \hline			
			\hline
		\end{tabular}
		\label{res_cgd}
	\end{center}
\end{table}

\subsection{Cancer Prevention Study}
\par Next, we reanalyze data on mammary cancer prevention by retinyl acetate (RA) in female rats \cite{gail80}. At day zero animals were injected with a carcinogen causing mammary cancer, followed by anti-cancer prevention for 60 days. After 60 days, the 48 animals
which remained tumour-free were randomly assigned to continued prophylaxis (active arm, $n=23$) or to placebo (control arm, $n=25$). Rats were palpated for tumours twice weekly, and observation ended 182 days after the initial carcinogen injection (122 days after the randomization). The outcome variable was time, measured in days, between the randomization at day 60 and the appearance of the first tumour and then the time elapsed between appearance of the following tumours up to censoring at the end of the study. The sizes of the clusters varied between 1 and 14 and the covariate reported in the original study, and hence readdressed in this paper, is the treatment arm indicator. Approximately $17\%$ of the observations were censored and, similarly to the CGD study, in this study sizes of the clusters have to be considered informative. Results, presented in Table \ref{res_ra}, show that the RA prophylaxis - compared to placebo - substantially reduces the hazard of mammary cancer by about half, which is also in agreement with the original paper. Estimated survival curves from the LPS method are shown in Figure \ref{fig:rl_curves} panel B. The estimated baseline survival curve indicates that the probability of remaining tumour-free decreases steadily over time, with a relatively higher risk in the early period. The treatment effect acts multiplicatively on this baseline, resulting in a consistently improved survival time in the treatment group as compared to the placebo.

\vspace{0.5cm}

\begin{table}[htb!]
    \begin{center}
	\caption{Results of the LPS and the CPH in the tumor prevention study. CrI, CI, and RA denote Credible Interval, Confidence Interval and Retinyl 
	Acetate, respectively.}
		\begin{tabular}{l | cccc}
			\hline
			\hline
            \multicolumn{1}{c |}{} & \multicolumn{2}{c}{LPS approach} & \multicolumn{2}{c}{CPH approach} \\
			Parameter & Estimate & $95\%$ CrI & Estimate & $95\%$ CI \\ 
			\hline			
			RA Treatment     & $-0.772$  & [$-1.171$; $-0.374$]   & $-0.745$ & [$-1.112$; $-0.378$]   \\			
			$\gamma$         & $\phantom{-}5.078$     &  $\phantom{-}[1.665;   \phantom{0}15.492$]     & $\phantom{-}7.492$   &  -- \\
            \hline			
			\hline
		\end{tabular}
		\label{res_ra}
	\end{center}
\end{table}

\subsection{Kidney Transplantation Study}
\par Finally, a study on kidney transplantation from deceased donors is considered \cite{Coll}. Kidneys donated by the same donor obviously share common (latent) characteristics. Therefore, it is plausible to assume that the survival times of two recipients who obtained organs from the same donor are correlated. This, again, leads to a hierarchical data structure, with intra-cluster correlation accounted for by a shared frailty. On the other hand, factors associated with a recipient are apparently important, like age or presence of underlying diseases. In the study considered here, 434 transplants using organs from 270 donors were analyzed, of whom 106 gave one kidney (which are essentially singleton clusters) and 164 donated both kidneys (clusters of size two). The cluster sizes are not informative. The outcome variable, the time to the graft rejection or death (in days), was modelled as function of the recipients age (in years) and presence of diabetes mellitus (1 if present and 0 otherwise). Approximately $84\%$ of the observations were right-censored.
\par Results, presented in Table \ref{res_kid}, show that age and presence of diabetes only slightly modify hazard of the graft rejection or death in kidney recipients which, again, stays in full agreement with the frequentist results. Also in this dataset, the estimated survival curves of the LPS model, which are presented for a 50-year-old subject (Figure \ref{fig:rl_curves}, panel C) overlap with the curve from the KM estimator. Note that the LPS curves lie entirely within the CIs of the KM curves.

\vspace{0.5cm}

\begin{table}[htb!]
    \begin{center}
	\caption{Results of the LPS and the CPH in the kidney transplantation study. CrI and CI denote Credible Interval and Confidence Interval, 
	respectively.}
		\begin{tabular}{l | cccc}
			\hline
			\hline
            \multicolumn{1}{c |}{} & \multicolumn{2}{c}{LPS approach} & \multicolumn{2}{c}{CPH approach} \\
			Parameter & Estimate & $95\%$ CrI & Estimate & $95\%$ CI \\ 
			\hline
            
            Age [yrs]             & $\phantom{-}0.019$    & [$-0.002; \phantom{-}0.039$]    & $\phantom{-}0.019$   & [$-0.001;  \phantom{-}0.039$] \\
            Diabetes              & $-0.165$ & [$-1.042; \phantom{-}0.711$]    & $-0.164$ & [$-1.028;  \phantom{-}0.699$] \\
            $\gamma$              & $\phantom{-}2.230$    & $\phantom{-}[0.141; \phantom{0}35.256]$      & $\phantom{-}2.967$ & -- \\
                        
			\hline			
			\hline
		\end{tabular}
		\label{res_kid}
	\end{center}
\end{table}

\begin{figure}[H]
	\centering
		\vspace{-1cm}
		\includegraphics[width=1\textwidth]{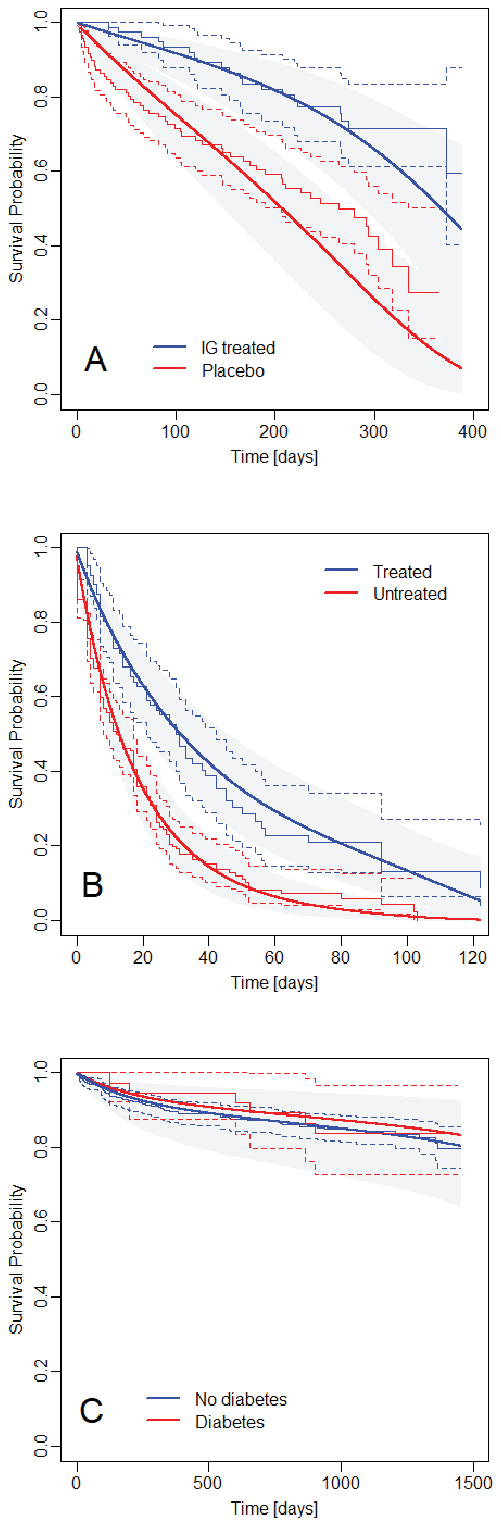}
		\caption{Estimated survival curves (blue and red solid curves) with their pointwise $95\%$ CrI (gray areas) obtained with the LPS model along with 
		the survival functions obtained with the Kaplan-Meier estimators and their $95\%$ CI (dashed lines). A, CGD study \cite{RN2284}; B, tumour 
		prevention study \cite{gail80}; C, kidney transplantation study \cite{Coll}.}
	\label{fig:rl_curves}
\end{figure}

\newpage

\section{Conclusion}
The main motivation to extend the existing LPS toolbox \cite{gressthes} to shared frailty survival comes from studies like those addressing the IG treatment in CGD \cite{RN2284}, experimental anti-tumour prevention therapy \cite{gail80} or survival time of multiorgan transplants from deceased donors \cite{Coll}. For proper treatment of such data, it is crucial to consider that the observed outcomes of interest (times-to-event) are not independent. Indeed, it is plausible to assume that times elapsing between incidences of a serious infection in a given patient are correlated, since some unobserved (latent) patient-specific characteristics may exist. It follows that those characteristics vary across patients, which leads to a hierarchical structure of the data with observations clustered within patients. In terms of survival analysis, the observations within a cluster (patient) share the same frailty.
\par Laplacian P-splines are particularly attractive for Bayesian inference, as they offer a much faster alternative as compared to classic MCMC procedures. E.g., a comparison with a sampling-based Metropolis-adjusted Langevin algorithm (MALA) revealed that the LPS method works approximately 10--40 times faster in the same simulation scenario \cite{Lewczuk_LPS}. Further, they permit to carry out inference in an entirely sampling-free framework, relying on analytically available gradient and Hessian formulas, speeding up the estimation process. In contrast to the CPH model, LPS procedure provides a smooth baseline survival curve.
\par This study is not without limitations. We believe that by looking at the data from the shared frailty perspective, we have accounted for the correlations across the multivariate outcomes for a given cluster, however, our model assumes that those correlations are time-invariant. This is a strong assumption that could be relaxed in future work. Furthermore, the shared frailty models only allow for a positive dependence between event times within the same cluster, which is not always realistic. Although the focus in this manuscript is on the shared Gamma frailty model, the Gamma frailty distribution is but one possible choice. Several other frailty distributions have been proposed in the literature \cite{wienke2010}. For example, the lognormal frailty distribution is often used in practice due to its close connection to normal random effects in mixed models. However, in that case the marginal likelihood has no closed-form expression and numerical methods are required. Despite the fact that the theoretical derivations within this manuscript are tailored to the Gamma frailty distribution, the expression in Eq.\ \eqref{general}, in terms of the derivatives of the Laplace transform, is more general. More specifically, if the derivatives of the Laplace transform of $U$ are available in closed form for a frailty distribution different from the Gamma distribution, the generalization of the LPS methodology proposed in this paper is straightforward. Finally, extensions of the proposed methodology to other types of censoring or the use of correlated frailty models provide avenues of further research.\\

\vspace{10cm}
\noindent The authors declare no conflict of interest. R script generating simulated datasets is available at \url{https://github.com/oswaldogressani/Frailty}. Other datasets considered in this paper are publicly available in the corresponding references.

\newpage

\bibliographystyle{unsrtnat}
\bibliography{BibliographyFrailty}

\newpage

\subsection*{Appendix}
\noindent \textbf{Gradient}

\noindent Let us first write the log-likelihood compactly as:

\vspace{-1cm}

\begin{eqnarray}
\ell(\bxi; \mathcal{D})=\sum_{i=1}^{\mathcal{I}} \Bigg(c_{\tilgamma} + \sum_{j=1}^{n_i}\delta_{ij}\left(\btheta^{\top}\boldsymbol{b}(t_{ij})+\bbeta^{\top} \boldsymbol{z}_{ij}\right)-(n_i \bar{\delta}_i+\exp(\tilgamma)) \log g_1(\btheta, \bbeta, \tilgamma) \Bigg), \nonumber 
\end{eqnarray}

\noindent with 

\vspace{-2cm}

\begin{eqnarray}
c_{\tilgamma}&:=&\exp(\tilgamma) \tilgamma + \log\Gamma(n_i \bar{\delta}_i+\exp(\tilgamma))- \log\Gamma(\exp(\tilgamma)), \nonumber \\
g_1(\btheta, \bbeta, \tilgamma)&:=&
\sum_{j=1}^{n_i}\Bigg\{\Bigg(\sum_{l=1}^{l(t_{ij})} \exp(\btheta^{\top}\boldsymbol{b}(s_l)) \Delta^* \Bigg) \exp(\bbeta^{\top} \boldsymbol{z}_{ij})\Bigg\} + \exp(\tilgamma). \nonumber 
\end{eqnarray}

\noindent The gradient is:

\vspace{-1cm}

\begin{eqnarray} \label{gradloglik}
\nabla_{\bxi} \ell(\bxi; \mathcal{D}):=\left(\frac{\partial \ell(\bxi; \mathcal{D})}{\partial \theta_1},\dots,\frac{\partial \ell(\bxi; \mathcal{D})}{\partial \theta_K},\frac{\partial \ell(\bxi; \mathcal{D})}{\partial \beta_1},\dots, \frac{\partial \ell(\bxi; \mathcal{D})}{\partial \beta_p},\frac{\partial \ell(\bxi; \mathcal{D})}{\partial \tilgamma}\right)^{\top}. \nonumber 
\end{eqnarray}

\noindent The derivative with respect to the $k$th B-spline coefficient (for $k=1,\dots,K$) is:

\vspace{-1cm}

\begin{eqnarray*}
\hspace{-0.5cm} \frac{\partial \ell(\bxi; \mathcal{D})}{\partial \theta_k}&=&\sum_{i=1}^{\mathcal{I}} \Bigg(\sum_{j=1}^{n_i} \delta_{ij} b_k(t_{ij})-(n_i \bar{\delta}_i+\exp(\tilgamma)) g_1^{-1}(\btheta, \bbeta, \tilgamma) \frac{\partial g_1(\btheta, \bbeta, \tilgamma)}{\partial \theta_k}\Bigg) \\
&=&\sum_{i=1}^{\mathcal{I}} \Bigg[ \sum_{j=1}^{n_i} \delta_{ij} b_k(t_{ij})-(n_i \bar{\delta}_i+\exp(\tilgamma)) g_1^{-1}(\btheta, \bbeta,\tilgamma) \\
&&\times \sum_{j=1}^{n_i}\Bigg\{\Bigg(\sum_{l=1}^{l(t_{ij})} \exp(\btheta^{\top}\boldsymbol{b}(s_l)) b_k(s_l) \Delta^* \Bigg) \exp(\bbeta^{\top} \boldsymbol{z}_{ij})\Bigg\} \Bigg]. \\
\end{eqnarray*}

\vspace{-0.5cm}

\noindent The derivative with respect to the $m$th regression coefficient (for $m=1,\dots,p$) is:

\vspace{-0.8cm}

\begin{eqnarray*}
\hspace{-0.5cm} \frac{\partial \ell(\bxi, \gamma; \mathcal{D})}{\partial \beta_m}&=&\sum_{i=1}^{\mathcal{I}} \Bigg(\sum_{j=1}^{n_i} \delta_{ij} z_{ijm}-(n_i \bar{\delta}_i+\exp(\tilgamma)) g_1^{-1}(\btheta, \bbeta,\tilgamma) \frac{\partial g_1(\btheta, \bbeta,\tilgamma)}{\partial \beta_m}\Bigg) \\
&=&\sum_{i=1}^{\mathcal{I}} \Bigg[\sum_{j=1}^{n_i} \delta_{ij} z_{ijm}-(n_i \bar{\delta}_i+\exp(\tilgamma)) g_1^{-1}(\btheta, \bbeta,\tilgamma) \\
&&\times \sum_{j=1}^{n_i}\Bigg\{\Bigg(\sum_{l=1}^{l(t_{ij})} \exp(\btheta^{\top}\boldsymbol{b}(s_l)) \Delta^* \Bigg) \exp(\bbeta^{\top} \boldsymbol{z}_{ij}) z_{ijm}\Bigg\} \Bigg]. \\
\end{eqnarray*}

\newpage 

\noindent The derivative with respect to the log-frailty term is:

\vspace{-0.8cm}

\begin{eqnarray}
\frac{\partial \ell(\bxi; \mathcal{D})}{\partial \tilgamma}&=&\sum_{i=1}^{\mathcal{I}} \Bigg( \frac{\partial c_{\tilgamma}}{\partial \tilgamma}-\Bigg[\frac{\partial (n_i \bar{\delta}_i+\exp(\tilgamma))}{\partial \tilgamma} \log g_1(\btheta, \bbeta, \tilgamma)+(n_i \bar{\delta}_i+\exp(\tilgamma)) \frac{\partial \log g_1(\btheta, \bbeta, \tilgamma)}{\partial \tilgamma}\Bigg]\Bigg). \nonumber 
\end{eqnarray}

\noindent Note that: 

\vspace{-1cm}

\begin{eqnarray}
\frac{\partial c_{\tilgamma}}{\partial \tilgamma}&=&\exp(\tilgamma) \tilgamma + \exp(\tilgamma) + \frac{\partial \log\Gamma(n_i \bar{\delta}_i+\exp(\tilgamma))}{\partial \tilgamma} - \frac{\partial \log\Gamma(\exp(\tilgamma))}{\partial \tilgamma} \nonumber \\
&=& \exp(\tilgamma)(1+\tilgamma)+\frac{\Gamma'(n_i \bar{\delta}_i+\exp(\tilgamma))}{\Gamma(n_i \bar{\delta}_i+\exp(\tilgamma))} \exp(\tilgamma)-\frac{\Gamma'(\exp(\tilgamma))}{\Gamma(\exp(\tilgamma))} \exp(\tilgamma) \nonumber \\
&=&\exp(\tilgamma)(1+\tilgamma) + \psi(n_i \bar{\delta}_i+\exp(\tilgamma)) \exp(\tilgamma) - \psi(\exp(\tilgamma)) \exp(\tilgamma) \nonumber \\
&=& \exp(\tilgamma)\left(1+\tilgamma+\psi(n_i \bar{\delta}_i+\exp(\tilgamma))- \psi(\exp(\tilgamma))\right), \nonumber 
\end{eqnarray}

\vspace{-0.2cm}

\noindent where $\psi(\cdot)$ is the digamma function. It follows that:

\vspace{-0.8cm}

\begin{eqnarray}
\frac{\partial \ell(\bxi; \mathcal{D})}{\partial \tilgamma}&=&\sum_{i=1}^{\mathcal{I}}\Bigg(\exp(\tilgamma)\Big(1+\tilgamma+\psi(n_i \bar{\delta}_i+\exp(\tilgamma))- \psi(\exp(\tilgamma))\Big)-\log g_1(\btheta, \bbeta, \tilgamma) \exp(\tilgamma) \nonumber \\
&&- (n_i \bar{\delta}_i+\exp(\tilgamma)) g_1^{-1}(\btheta, \bbeta, \tilgamma) \exp(\tilgamma) \Bigg) \nonumber \\
&=& \exp(\tilgamma) \sum_{i=1}^{\mathcal{I}}\Bigg(1+\tilgamma+\psi(n_i \bar{\delta}_i+\exp(\tilgamma))- \psi(\exp(\tilgamma)) \nonumber \\
&&- \log g_1(\btheta, \bbeta, \tilgamma) - (n_i \bar{\delta}_i+\exp(\tilgamma)) g_1^{-1}(\btheta, \bbeta, \tilgamma)\Bigg). \nonumber  
\end{eqnarray}

\noindent \textbf{Hessian}

\noindent To compute the Hessian, let us first define:

\vspace{-0.8cm}

\begin{eqnarray}
g_2(\btheta, \bbeta)&:=&\sum_{j=1}^{n_i}\Bigg\{\Bigg(\sum_{l=1}^{l(t_{ij})} \exp(\btheta^{\top}\boldsymbol{b}(s_l)) b_k(s_l) \Delta^* \Bigg) \exp(\bbeta^{\top} \boldsymbol{z}_{ij})\Bigg\}, \nonumber \\
g_3(\btheta, \bbeta)&:=&\sum_{j=1}^{n_i}\Bigg\{\Bigg(\sum_{l=1}^{l(t_{ij})} \exp(\btheta^{\top}\boldsymbol{b}(s_l)) \Delta^* \Bigg) \exp(\bbeta^{\top} \boldsymbol{z}_{ij}) z_{ijm}\Bigg\}, \nonumber \\
g_4(\btheta, \bbeta, \tilgamma)&:=&1+\tilgamma+\psi(n_i \bar{\delta}_i+\exp(\tilgamma))- \psi(\exp(\tilgamma))-\log g_1(\btheta, \bbeta, \tilgamma) - (n_i \bar{\delta}_i+\exp(\tilgamma)) g_1^{-1}(\btheta, \bbeta, \tilgamma). \nonumber 
\end{eqnarray}

\noindent This allows to write the derivative with respect to the $k$th B-spline coefficient compactly:

\vspace{-0.8cm}

\begin{eqnarray}
\frac{\partial \ell(\bxi; \mathcal{D})}{\partial \theta_k} = \sum_{i=1}^{\mathcal{I}} \Bigg(\sum_{j=1}^{n_i} \delta_{ij} b_k(t_{ij})-(n_i \bar{\delta}_i+\exp(\tilgamma)) \frac{g_2(\btheta, \bbeta)}{g_1(\btheta, \bbeta, \tilgamma)}\Bigg). \nonumber
\end{eqnarray}

\newpage

\noindent The second-order partial derivatives with respect to the spline components are given by:

\vspace{-0.8cm}

\begin{eqnarray*}
\frac{\partial^2 \ell(\bxi; \mathcal{D})}{\partial \theta_k \partial \theta_{k'}}=-\sum_{i=1}^{\mathcal{I}} \Bigg((n_i \bar{\delta}_i+\exp(\tilgamma)) \Big(g_1^2(\btheta, \bbeta, \tilgamma)\Big)^{-1} \Bigg(\frac{\partial g_2(\btheta, \bbeta)}{\partial \theta_{k'}} g_1(\btheta, \bbeta,\tilgamma)-g_2(\btheta, \bbeta) \frac{\partial g_1(\btheta, \bbeta,\tilgamma)}{\partial \theta_{k'}}\Bigg) \Bigg),
\end{eqnarray*}

\noindent where 

\vspace{-1.3cm}

\begin{eqnarray*}
&&\frac{\partial g_1(\btheta, \bbeta, \tilgamma)}{\partial \theta_{k'}} = \sum_{j=1}^{n_i}\Bigg\{\Bigg(\sum_{l=1}^{l(t_{ij})} \exp(\btheta^{\top}\boldsymbol{b}(s_l)) b_{k'}(s_l) \Delta^* \Bigg) \exp(\bbeta^{\top} \boldsymbol{z}_{ij})\Bigg\},\\
&&\frac{\partial g_2(\btheta, \bbeta)}{\partial \theta_{k'}} = \sum_{j=1}^{n_i}\Bigg\{\Bigg(\sum_{l=1}^{l(t_{ij})} \exp(\btheta^{\top}\boldsymbol{b}(s_l))b_k(s_l) b_{k'}(s_l) \Delta^* \Bigg) \exp(\bbeta^{\top} \boldsymbol{z}_{ij})\Bigg\}, \\
\end{eqnarray*}

\vspace{-0.8cm} 

\noindent for $k=1,\dots,K$ and $k'=1,\dots,K$.

\noindent The cross partial derivatives between the spline components and the regression coefficients are given by:

\vspace{-1cm}

\begin{eqnarray}
\frac{\partial^2 \ell(\bxi; \mathcal{D})}{\partial \theta_k \partial \beta_m}
&=&-\sum_{i=1}^{\mathcal{I}} \Bigg((n_i \bar{\delta}_i+\exp(\tilgamma)) \Big(g_1^2(\btheta, \bbeta, \tilgamma)\Big)^{-1} \Bigg(\frac{\partial g_2(\btheta, \bbeta)}{\partial \beta_m} g_1(\btheta, \bbeta, \tilgamma)-g_2(\btheta, \bbeta) \frac{\partial g_1(\btheta, \bbeta, \tilgamma)}{\partial \beta_m}\Bigg) \Bigg), \nonumber 
\end{eqnarray}

\noindent where 

\vspace{-1.3cm}

\begin{eqnarray*}
&&\frac{\partial g_1(\btheta, \bbeta, \tilgamma)}{\partial \beta_m} = g_3(\btheta, \bbeta), \\
&&\frac{\partial g_2(\btheta, \bbeta)}{\partial \beta_m} = \frac{\partial g_3(\btheta, \bbeta)}{\partial \theta_k} =  \sum_{j=1}^{n_i}\Bigg\{\Bigg(\sum_{l=1}^{l(t_{ij})} \exp(\btheta^{\top}\boldsymbol{b}(s_l)) b_k(s_l) \Delta^* \Bigg) \exp(\bbeta^{\top} \boldsymbol{z}_{ij}) z_{ijm}\Bigg\}, \\
\end{eqnarray*}

\vspace{-0.8cm}

\noindent for $k=1,\dots,K$ and $m=1,\dots,p$.

\noindent The cross partial derivatives between the spline components and the log-frailty term are:

\vspace{-1cm}

\begin{eqnarray}
\frac{\partial^2 \ell(\bxi; \mathcal{D})}{\partial \theta_k \partial \tilgamma}&=&-\sum_{i=1}^{\mathcal{I}}\Bigg(\frac{\partial (n_i \bar{\delta}_i+\exp(\tilgamma))}{\partial \tilgamma}\frac{g_2(\btheta, \bbeta)}{g_1(\btheta, \bbeta, \tilgamma)} + (n_i \bar{\delta}_i+\exp(\tilgamma)) g_2(\btheta, \bbeta) \frac{\partial g_1^{-1}(\btheta, \bbeta, \tilgamma)}{\partial \tilgamma}\Bigg) \nonumber \\
&=&-\sum_{i=1}^{\mathcal{I}}\Bigg(\exp(\tilgamma) \frac{g_2(\btheta, \bbeta)}{g_1(\btheta, \bbeta, \tilgamma)}-(n_i \bar{\delta}_i+\exp(\tilgamma)) \frac{g_2(\btheta, \bbeta)}{g_1^{2}(\btheta, \bbeta, \tilgamma)} \exp(\tilgamma)\Bigg) \nonumber \\
&=&-\exp(\tilgamma)\sum_{i=1}^{\mathcal{I}}\Bigg(\frac{g_2(\btheta, \bbeta)}{g_1(\btheta, \bbeta, \tilgamma)}-(n_i \bar{\delta}_i+\exp(\tilgamma)) \frac{g_2(\btheta, \bbeta)}{g_1^{2}(\btheta, \bbeta, \tilgamma)}\Bigg). \nonumber
\end{eqnarray}

\noindent A direct consequence of the above partial derivatives is that $\frac{\partial^2 \ell(\bxi; \mathcal{D})}{\partial \beta_m \partial \theta_k}=\frac{\partial^2 \ell(\bxi; \mathcal{D})}{\partial \theta_k \partial \beta_m}$ and $\frac{\partial^2 \ell(\bxi; \mathcal{D})}{\partial \tilgamma \partial \theta_k}=\frac{\partial^2 \ell(\bxi; \mathcal{D})}{\partial \theta_k \partial \tilgamma}$. Using $g_3(\btheta, \bbeta)$, the derivative with respect to the $m$th regression coefficient is:

\vspace{-0.4cm}

\begin{eqnarray}
\frac{\partial \ell(\bxi; \mathcal{D})}{\partial \beta_m}=\sum_{i=1}^{\mathcal{I}} \Bigg(\sum_{j=1}^{n_i} \delta_{ij} z_{ijm}-(n_i \bar{\delta}_i+\exp(\tilgamma)) \frac{g_3(\btheta, \bbeta)}{g_1(\btheta, \bbeta,\tilgamma)} \Bigg). \nonumber
\end{eqnarray}

\vspace{-0.1cm}

\noindent The second-order partial derivatives with respect to the regression coefficients are given by:

\vspace{-0.8cm}

\begin{eqnarray}
\frac{\partial^2 \ell(\bxi; \mathcal{D})}{\partial \beta_m \partial \beta_{m'}}=-\sum_{i=1}^{\mathcal{I}} \Bigg((n_i \bar{\delta}_i+\exp(\tilgamma)) \Big(g_1^2(\btheta, \bbeta,\tilgamma)\Big)^{-1} \Bigg(\frac{\partial g_3(\btheta, \bbeta)}{\partial \beta_{m'}} g_1(\btheta, \bbeta,\tilgamma)-g_3(\btheta, \bbeta) \frac{\partial g_1(\btheta, \bbeta,\tilgamma)}{\partial \beta_{m'}}\Bigg) \Bigg), \nonumber 
\end{eqnarray}

\noindent where 

\vspace{-1.3cm}

\begin{eqnarray*}
&&\frac{\partial g_1(\btheta, \bbeta,\tilgamma)}{\partial \beta_{m'}} = \sum_{j=1}^{n_i}\Bigg\{\Bigg(\sum_{l=1}^{l(t_{ij})} \exp(\btheta^{\top}\boldsymbol{b}(s_l)) \Delta^* \Bigg) \exp(\bbeta^{\top} \boldsymbol{z}_{ij}) z_{ij{m'}}\Bigg\}, \\
&&\frac{\partial g_3(\btheta, \bbeta)}{\partial \beta_{m'}} = \sum_{j=1}^{n_i}\Bigg\{\Bigg(\sum_{l=1}^{l(t_{ij})} \exp(\btheta^{\top}\boldsymbol{b}(s_l)) \Delta^* \Bigg) \exp(\bbeta^{\top} \boldsymbol{z}_{ij}) z_{ijm} z_{ij{m'}}\Bigg\}, \\
\end{eqnarray*}

\vspace{-0.8cm}

\noindent for $m=1,\dots,p$ and $m'=1,\dots,p$.

\noindent The cross partial derivatives between the regression coefficients and the log-frailty term are: 

\vspace{-0.7cm}

\begin{eqnarray}
\frac{\partial^2 \ell(\bxi; \mathcal{D})}{\partial \beta_m \partial \tilgamma}&=&-\sum_{i=1}^{\mathcal{I}}\Bigg(\frac{\partial (n_i \bar{\delta}_i+\exp(\tilgamma))}{\partial \tilgamma} \frac{g_3(\btheta, \bbeta)}{g_1(\btheta, \bbeta,\tilgamma)} + (n_i \bar{\delta}_i+\exp(\tilgamma)) g_3(\btheta, \bbeta) \frac{\partial g_1^{-1}(\btheta, \bbeta,\tilgamma)}{\partial \tilgamma}\Bigg) \nonumber \\
&=&-\sum_{i=1}^{\mathcal{I}}\Bigg(\exp(\tilgamma) \frac{g_3(\btheta, \bbeta)}{g_1(\btheta, \bbeta, \tilgamma)}-(n_i \bar{\delta}_i+\exp(\tilgamma)) \frac{g_3(\btheta, \bbeta)}{g_1^{2}(\btheta, \bbeta, \tilgamma)} \exp(\tilgamma)\Bigg) \nonumber \\
&=&-\exp(\tilgamma)\sum_{i=1}^{\mathcal{I}}\Bigg(\frac{g_3(\btheta, \bbeta)}{g_1(\btheta, \bbeta, \tilgamma)}-(n_i \bar{\delta}_i+\exp(\tilgamma)) \frac{g_3(\btheta, \bbeta)}{g_1^{2}(\btheta, \bbeta, \tilgamma)}\Bigg) \nonumber \\
&=& \frac{\partial^2 \ell(\bxi; \mathcal{D})}{\partial \tilgamma \partial \beta_m},\ \text{for } m=1,\dots,p. \nonumber
\end{eqnarray}

\noindent To obtain the second-order partial derivative of the log-frailty term, recall that:

\vspace{-1cm}

\begin{eqnarray}
\hspace{-0.9cm} \frac{\partial \ell(\bxi; \mathcal{D})}{\partial \tilgamma}&=&\exp(\tilgamma) \sum_{i=1}^{\mathcal{I}}\Bigg(1+\tilgamma+\psi(n_i \bar{\delta}_i+\exp(\tilgamma))- \psi(\exp(\tilgamma)) \nonumber \\
&& -\log g_1(\btheta, \bbeta, \tilgamma) - (n_i \bar{\delta}_i+\exp(\tilgamma)) g_1^{-1}(\btheta, \bbeta, \tilgamma)\Bigg) \nonumber \\
&=& \exp(\tilgamma) \sum_{i=1}^{\mathcal{I}} g_4(\btheta, \bbeta, \tilgamma). \nonumber 
\end{eqnarray}

\noindent It follows that:

\begin{eqnarray}
\hspace{-0.9cm} \frac{\partial^2 \ell(\bxi; \mathcal{D})}{\partial \tilgamma^2}&=&\exp(\tilgamma) \sum_{i=1}^{\mathcal{I}} g_4(\btheta, \bbeta, \tilgamma) + \exp(\tilgamma) \sum_{i=1}^{\mathcal{I}} \frac{\partial{g_4(\btheta, \bbeta, \tilgamma)}}{\partial \tilgamma}, \nonumber 
\end{eqnarray}

\vspace{3cm}

\noindent where

\vspace{-0.5cm}

\begin{eqnarray}
\frac{\partial{g_4(\btheta, \bbeta, \tilgamma)}}{\partial \tilgamma}&=&1+\frac{\partial \psi(n_i \bar{\delta}_i+\exp(\tilgamma))}{\partial \tilgamma} - \frac{\partial \psi(\exp(\tilgamma))}{\partial \tilgamma}-\frac{\partial \log g_1(\btheta, \bbeta, \tilgamma)}{\partial \tilgamma} \nonumber \\
&&-\Bigg(\frac{\partial (n_i \bar{\delta}_i+\exp(\tilgamma))}{\partial \tilgamma}g_1^{-1}(\btheta, \bbeta, \tilgamma)+(n_i \bar{\delta}_i+\exp(\tilgamma)) \frac{\partial g_1^{-1}(\btheta, \bbeta, \tilgamma)}{\partial \tilgamma}\Bigg) \nonumber \\
&=& 1 + \psi^{(1)}(n_i \bar{\delta}_i+\exp(\tilgamma)) \exp(\tilgamma)-\psi^{(1)}(\exp(\tilgamma)) \exp(\tilgamma)-g_1^{-1}(\btheta, \bbeta, \tilgamma) \exp(\tilgamma) \nonumber \\
&&-g_1^{-1}(\btheta, \bbeta, \tilgamma)\exp(\tilgamma)+\frac{(n_i \bar{\delta}_i+\exp(\tilgamma))}{g_1^{2}(\btheta, \bbeta, \tilgamma)} \exp(\tilgamma) \nonumber \\
&=& 1 + \exp(\tilgamma) \Bigg(\psi^{(1)}(n_i \bar{\delta}_i+\exp(\tilgamma)) -\psi^{(1)}(\exp(\tilgamma))-2g_1^{-1}(\btheta, \bbeta, \tilgamma)+\frac{(n_i \bar{\delta}_i+\exp(\tilgamma))}{g_1^{2}(\btheta, \bbeta, \tilgamma)}\Bigg), \nonumber 
\end{eqnarray}

\noindent where $\psi^{(1)}(\cdot)$ is the trigamma function. All the above results can be used to compute the Hessian matrix of the log-likelihood:

\begin{eqnarray}
 \nabla_{\bxi}^2 \ell(\bxi; \mathcal{D}) = \begin{pmatrix}
 \frac{\partial^2 \ell(\bxi; \mathcal{D})}{\partial \btheta \partial \btheta^{\top}} & \frac{\partial^2 \ell(\bxi; \mathcal{D})}{\partial \btheta \partial \bbeta^{\top}} &  \frac{\partial^2\ell(\bxi; \mathcal{D})}{\partial \btheta \partial \tilgamma} \\
 \frac{\partial^2 \ell(\bxi, \gamma; \mathcal{D})}{\partial \bbeta \partial \btheta^{\top}} & 
 \frac{\partial^2 \ell(\bxi, \gamma; \mathcal{D})}{\partial \bbeta \partial \bbeta^{\top}} & \frac{\partial^2\ell(\bxi; \mathcal{D})}{\partial \bbeta \partial \tilgamma} \\
 \frac{\partial^2\ell(\bxi; \mathcal{D})}{\partial \tilgamma \partial \btheta^{\top}} & \frac{\partial^2\ell(\bxi; \mathcal{D})}{\partial \tilgamma \partial \bbeta^{\top}} & \frac{\partial^2\ell(\bxi; \mathcal{D})}{\partial \tilgamma^2}
 \end{pmatrix}. \nonumber 
\end{eqnarray}

\end{document}